\documentclass[modern]{aastex631}

\usepackage[encapsulated]{CJK}

\newcommand\logg{$\log~g$}
\newcommand\msun{M$_\odot$}

\accepted{22 Nov 2023}

\shorttitle{JWST Spectra of Distant T Dwarfs}
\shortauthors{Burgasser et al.}
\graphicspath{{./}{}}
\begin{document}

\title{UNCOVER: JWST Spectroscopy of Three Cold Brown Dwarfs at Kiloparsec-scale Distances}

\correspondingauthor{Adam Burgasser}
\email{aburgasser@ucsd.edu}

\author[0000-0002-6523-9536]{Adam J.\ Burgasser}
\affiliation{Department of Astronomy \& Astrophysics, UC San Diego, La Jolla, CA 92093, USA}

%%%%%%% UNCOVER TEAM PLEASE ADD IN YOUR AUTHOR INFO HERE %%%%%%%

% PIS
\author[0000-0001-5063-8254]{Rachel Bezanson}
\affiliation{Department of Physics and Astronomy and PITT PACC, University of Pittsburgh, Pittsburgh, PA 15260, USA}

\author[0000-0002-2057-5376]{Ivo Labbe}
\affiliation{Centre for Astrophysics and Supercomputing, Swinburne University of Technology, Melbourne, VIC 3122, Australia}

% BUILDERS
\author[0000-0003-2680-005X]{Gabriel Brammer}
\affiliation{Cosmic Dawn Center (DAWN), Niels Bohr Institute, University of Copenhagen, Jagtvej 128, K{\o}benhavn N, DK-2200, Denmark}

\author[0000-0002-7031-2865]{Sam E. Cutler}
\affiliation{Department of Astronomy, University of Massachusetts, Amherst, MA 01003, USA}

\author[0000-0001-6278-032X]{Lukas J. Furtak}
\affiliation{Physics Department, Ben-Gurion University of the Negev, P.O. Box 653, Be’er-Sheva 84105, Israel}

\author[0000-0002-5612-3427]{Jenny E. Greene}
\affiliation{Department of Astrophysical Sciences, Princeton University, 4 Ivy Lane, Princeton, NJ 08544}

\author[0000-0003-0398-639X]{Roman Gerasimov}
\affiliation{Department of Physics, UC San Diego, La Jolla, CA 92093, USA}
\affiliation{Department of Physics \& Astronomy, University of Notre Dame, Notre Dame, IN 46556, USA}

\author[0000-0001-6755-1315]{Joel Leja}
\affiliation{Department of Astronomy \& Astrophysics, The Pennsylvania State University, University Park, PA 16802, USA}
\affiliation{Institute for Computational \& Data Sciences, The Pennsylvania State University, University Park, PA 16802, USA}
\affiliation{Institute for Gravitation and the Cosmos, The Pennsylvania State University, University Park, PA 16802, USA}

\author[0000-0002-9651-5716]{Richard Pan}\affiliation{Department of Physics and Astronomy, Tufts University, 574 Boston Ave., Medford, MA 02155, USA}

\author[0000-0002-0108-4176]{Sedona H. Price}
\affiliation{Department of Physics and Astronomy and PITT PACC, University of Pittsburgh, Pittsburgh, PA 15260, USA}

\author[0000-0001-9269-5046]{Bingjie Wang (\begin{CJK*}{UTF8}{gbsn}王冰洁\ignorespacesafterend\end{CJK*})}
\affiliation{Department of Astronomy \& Astrophysics, The Pennsylvania State University, University Park, PA 16802, USA}
\affiliation{Institute for Computational \& Data Sciences, The Pennsylvania State University, University Park, PA 16802, USA}
\affiliation{Institute for Gravitation and the Cosmos, The Pennsylvania State University, University Park, PA 16802, USA}

\author[0000-0003-1614-196X]{John R. Weaver}
\affiliation{Department of Astronomy, University of Massachusetts, Amherst, MA 01003, USA}

\author[0000-0001-7160-3632]{Katherine E. Whitaker}
\affiliation{Department of Astronomy, University of Massachusetts, Amherst, MA 01003, USA}
\affiliation{Cosmic Dawn Center (DAWN), Denmark} 

\author[0000-0001-7201-5066]{Seiji Fujimoto}\altaffiliation{Hubble Fellow}
\affiliation{
Department of Astronomy, The University of Texas at Austin, Austin, TX 78712, USA}

\author[0000-0002-5588-9156]{Vasily Kokorev}
\affiliation{Kapteyn Astronomical Institute, University of Groningen, 9700 AV Groningen, The Netherlands}

\author[0000-0001-8460-1564]{Pratika Dayal}
\affiliation{Kapteyn Astronomical Institute, University of Groningen, 9700 AV Groningen, The Netherlands}

\author[0000-0003-2804-0648 ]{Themiya Nanayakkara}
\affiliation{Centre for Astrophysics and Supercomputing, Swinburne University of Technology, PO Box 218, Hawthorn, VIC 3122, Australia}

\author[0000-0003-2919-7495]{Christina C. Williams}
\affiliation{NSF’s National Optical-Infrared Astronomy Research Laboratory, 950 N. Cherry Avenue, Tucson, AZ 85719, USA}

\author[0000-0001-9002-3502]{Danilo Marchesini}
\affiliation{Department of Physics \& Astronomy, Tufts University, MA 02155, USA}

\author[0000-0002-0350-4488]{Adi Zitrin}
\affiliation{Physics Department, Ben-Gurion University of the Negev, P.O. Box 653, Be'er-Sheva 84105, Israel}

\author[0000-0002-8282-9888]{Pieter van Dokkum}
\altaffiliation{Astronomy Department, Yale University, 52 Hillhouse Ave, New Haven, CT 06511, USA}

\begin{abstract}
We report JWST/NIRSpec spectra of three distant T-type brown dwarfs identified in the Ultradeep NIRSpec and NIRCam ObserVations before the Epoch of Reionization (UNCOVER) survey of the Abell 2744 lensing field. One source was previously reported as a candidate T dwarf on the basis of NIRCam photometry, while two sources were initially identified as candidate active galactic nuclei. Low-resolution 1--5~$\micron$ spectra confirm the presence of molecular features consistent with T dwarf atmospheres, and comparison to spectral standards infers classifications of sdT1, T6, and T8--T9. The warmest source, UNCOVER-BD-1, shows evidence of subsolar metallicity, and atmosphere model fits indicates T$_{eff}$ = 1300~K and [M/H] $\sim$ $-$1.0, making this one of the few spectroscopically-confirmed T subdwarfs known. The coldest source, UNCOVER-BD-3, is near the T/Y dwarf boundary with T$_{eff}$ = 550~K, and our analysis indicates the presence of PH$_3$ in the 3--5~$\micron$ region, favored over CO$_2$ and a possible indicator of subsolar metallicity. We estimate distances of 0.9--4.5~kpc from the Galactic midplane, making these the most distant brown dwarfs with spectroscopic confirmation. Population simulations indicate high probabilities of membership in the Galactic thick disk {for two of these brown dwarfs, and potential halo membership for UNCOVER-BD-1}.
Our simulations indicate that there are {approximately 5} T dwarfs and {1--2} L dwarfs in the Abell 2744 field down to F444W = 30 AB mag, roughly {one-third of which are thick disk} members. These results highlight the utility of deep JWST/NIRSpec spectroscopy for identifying and characterizing the oldest metal-poor brown dwarfs in the Milky Way.
\end{abstract}


\section{Introduction} \label{sec:intro}

Brown dwarfs are stellar objects with masses below the $\sim$0.075~{\msun} threshold for sustained core hydrogen fusion \citep{1962AJ.....67S.579K,1963ApJ...137.1121K,1963PThPh..30..460H}.
Supported by electron degeneracy pressure, these compact, hydrogen-rich objects radiate their initial heat of formation and continuously cool over time. The relatively high abundance of brown dwarfs in the immediate vicinity of the Sun ($\gtrsim$20\% of stars; \citealt{2021ApJS..253....7K,2021A&A...650A.201R}), their lack of chemical processing, and their time-dependent luminosities make them useful probes of the ages and chemical evolution of various Milky Way populations \citep{1998ApJ...499L.199S,2022ApJ...930...24G}. However, the intrinsic faintness of cool brown dwarfs largely limits their detection to the immediate Solar Neighborhood ($d \lesssim$ 100~pc). 

The unprecedented sensitivity of JWST at near-infrared wavelengths where brown dwarfs are brightest enables detection of brown dwarfs deep into the Galactic halo \citep{2016AJ....151...92R,2022ApJ...934...73A}. Multiple brown dwarf candidates have already been identified in deep multi-band imaging surveys \citep{2023ApJ...942L..29N,2023ApJ...947L..25G,2023MNRAS.523.4534W,2023arXiv230903250H,2023arXiv230905835H}.
However, confirmation and physical characterization of these distant brown dwarfs, including accurate measurement of temperatures, surface gravities, and compositions, requires spectroscopy. 

The JWST Cycle 1 Treasury program Ultradeep NIRSpec and NIRCam
ObserVations before the Epoch of Reionization (UNCOVER; 
 \citealt{2022arXiv221204026B}) aims to identify 
% the highest redshift systems that are lensed by the massive z=0.308 galaxy cluster Abell 2744. 
some of the highest redshift and faintest Epoch of Reionization (EoR) systems that are lensed by the massive z=0.308 galaxy cluster Abell 2744. 
UNCOVER combines deep JWST NIRCam imaging over $\sim45$ arcmin$^2$ in seven filters
(F115W, F150W, F200W, F277W, F356W, F410M, F444W) down to $\sim30$ AB mag \citep{2023arXiv230102671W} with NIRSpec PRISM 0.6--5~$\micron$ low-resolution spectroscopy.
% [HIGHLIGHT SOME RESULTS]
% To date, UNCOVER has photometrically identified and characterized $\sim50,000$ galaxies in the Abell 2744 field, and has spectroscopically confirmed 
% % {\color{red}{multiple}} galaxies at $z\gtrsim10$ \citep[][{\color{red}{also add Seiji's, depending on submission timelines}}]{2023arXiv230803745W}.
% galaxies out to $z\gtrsim10$ (e.g., \citealt{2023arXiv230803745W}).
% Deep NIRSpec spectroscopy has also
% characterized the hosts and supermassive black holes of 
% multiple high-redshift 
% active galactic nuclei (AGN; e.g., \citealt{2023arXiv230802750G, 2023arXiv230805735F}), 
% % a $z\sim10$ X-ray luminous active galactic nuclei (AGN; \citealt{2023arXiv230802750G}), 
% and performed the first spectroscopic analysis of low luminosity EoR galaxies \citep{2023arXiv230808540A}. 
%%%%
%%%%
To date, UNCOVER has photometrically identified and characterized $\sim50,000$ galaxies in the Abell 2744 field, and has spectroscopically confirmed galaxies out to $z\gtrsim10$ (e.g., \citealt{2023arXiv230803745W}), 
characterized the hosts and supermassive black holes of multiple high-redshift active galactic nuclei (AGN; e.g., \citealt{2023arXiv230802750G, 2023arXiv230805735F}), 
and performed the first spectroscopic analysis of low luminosity EoR galaxies \citep{2023arXiv230808540A}. 

% To date, UNCOVER has identified and confirmed galaxies at $z\gtrsim12$ 
% \citep{2023arXiv230803745W} ({\color{red}{extend down to 9-10 if we include Seiji's paper}}, confirmed the redshift of a $z\sim10$ X-ray luminous AGN 
% \citep{2023arXiv230802750G}, 
% characterized a faint, reddened AGN at $z=7$ with a $\sim10^{7}M_{\odot}$ supermassive black hole, 

In this article, we utilize the deep imaging and spectroscopy of UNCOVER to identify and spectroscopically characterize three T-type brown dwarfs. 
Section~2 summarizes the identification and observations of these sources.
Section~3 describes our analysis of JWST/NIRSpec spectra, including classification, model fitting, and estimation of the distances by scaling model surface fluxes.
Section~4 presents population simulations that justifies the detection of three T dwarfs in the narrow Abell 2744 field, and provides statistical constraints on the Galactic population membership.
Section 5 summarizes our results.
{We note that these sources and analysis of the data presented here have also been reported contemporaneously by \citet{2023arXiv230810900L}.}

\section{Identification and Observations}

Catalog designations, coordinates, and AB magnitudes from \citet{2023arXiv230102671W} of the three sources presented here are listed in Table~\ref{tab:properties}. Two of the sources were initially identified as candidate high-redshift AGN in UNCOVER NIRCam photometry \citep{2023arXiv230607320L}. The third source, 39243, was previously identified as the candidate T dwarf GLASS-BD-1 {(aka ``Nonino's Dwarf'')} based on multi-band NIRCam photometry from the Grism Lens-Amplified Survey from Space (GLASS) survey \citep{2023ApJ...942L..29N,2023ApJ...947L..25G}. All three sources were included in the first UNCOVER NIRSpec campaign.

% [DETAILS ON SPECTRAL OBSERVATIONS]
NIRSpec/PRISM observations were obtained on 
31 July -- 2 August 2023 (UT), split over seven Multi-shutter array (MSA) masks. 
All observations were taken with a 2-POINT-WITH-NIRCam-SIZE2 dither pattern and a three shutter slitlet nod pattern at an angle of $\mathrm{V3PA\sim 266^{\circ}}$. 
Sources 32265 and 33437 were observed on MSAs 3, 5, 6, and 7, using the NRSIRS2RAPID readout pattern for MSA 3 and NRSIRS2 for the rest, for a total of 14.3~hours.
Source ID 39243 was observed on MSA 4 with readout pattern NRSIRS2 for a total of 4.4~hours.
%%%
% Mask 3 and masks 4-7 used the NRSIRS2RAPID and NRSIRS2 readout patterns, respectively. 
% {\color{red}{(32265 and 33437 were observed on masks 3,5,6,7 and 39243 on mask 4).}}
% 32265: MSA3,5,6,7:  14.5h
% 33437: MSA3,5,6,7: 14.5h
% 39243: MSA4: 4.4h
% %%%
%%%%
Full observational details will be presented in Price et al. (2023, in prep). 
%%%

The spectra were reduced using \texttt{msaexp} (v0.6.10; \citealt{Brammer2022}), 
starting from the level 2 data products downloaded from the Mikulski Archive for Space Telescopes (MAST)\footnote{{\dataset[10.17909/8k5c-xr27]{\doi{10.17909/8k5c-xr27}}}},
{reduced using version 1.11.3 of the JWST Calibration pipeline \citep{bushouse_2023_8157276}.}
For each slit mask, \texttt{msaexp} corrects $1/f$-noise, finds and masks cosmic ray ``snowballs,'' and removes the bias in each individual frame. It then applies a world coordinate system, identifies each slit, and applies flat-fielding and photometric corrections. 
The 2D slits are combined and drizzled onto a common grid, a local background subtraction is applied, and spectra are then optimally extracted \citep{1986PASP...98..609H}. 
For the early spectroscopic reduction presented here (internal v0.3), flux calibration was modeled as a first-order polynomial  
determined by convolving the individual mask 1D spectra with the broad/medium band filters, and comparing to the total photometry \citep{2023arXiv230102671W}. 

\begin{figure}[t]
\plotone{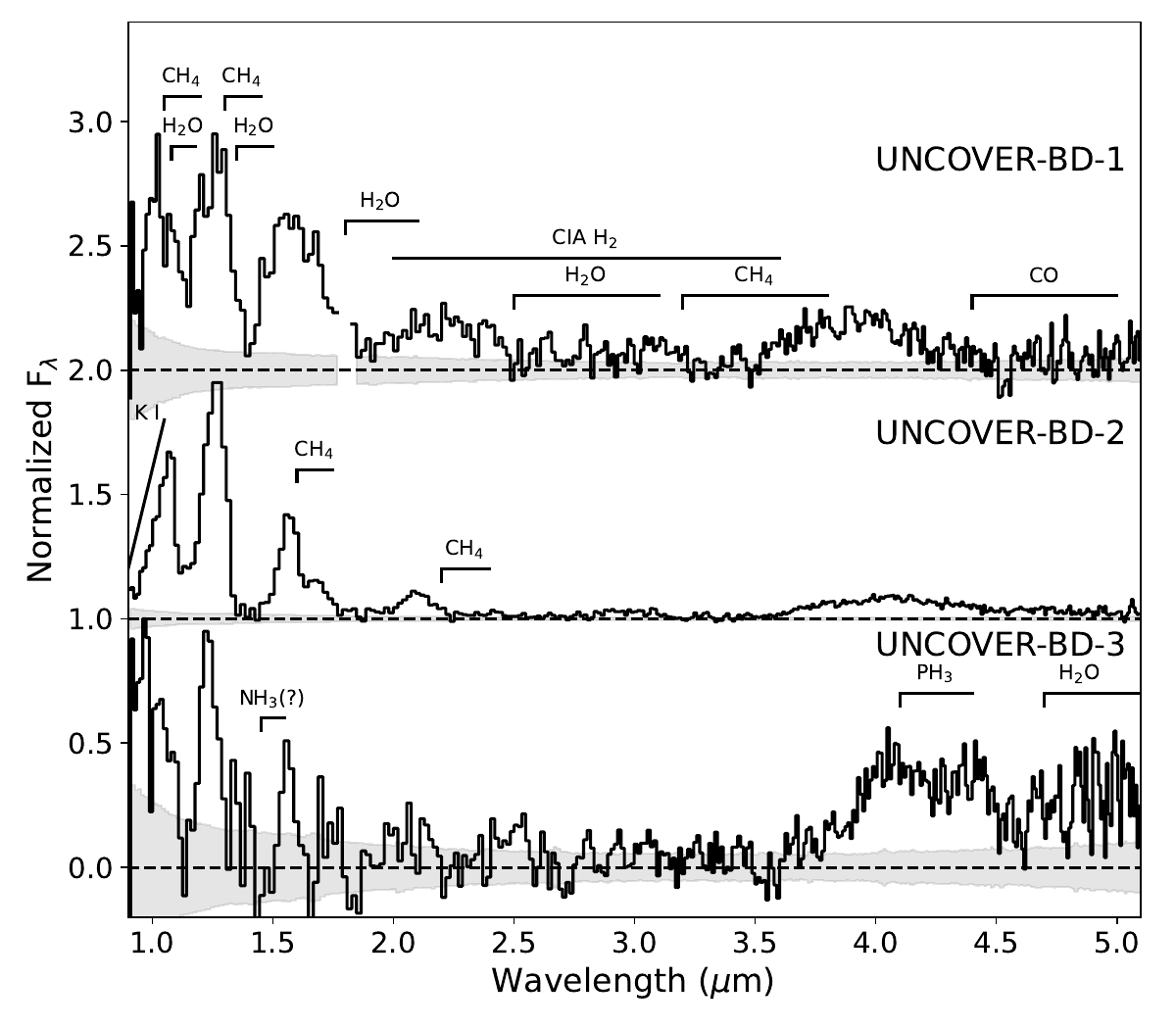}
\caption{JWST/NIRSpec PRISM spectra of the three brown dwarf candidates observed by UNCOVER in $F_{\lambda}$ units (black lines with grey shaded uncertainties), 
{normalized in the 1.1--1.3~$\mu$m region to a maximum value of 0.95.}
Each spectrum is offset by a constant (dashed lines), and major spectral features in the 1--5~$\micron$ region are labeled.
{A bad pixel at 1.8~$\mu$m has been masked out in the spectrum of UNCOVER-BD-1.}
\label{fig:spectra}}
\end{figure}

Figure~\ref{fig:spectra} shows the calibrated JWST/NIRSpec spectra over the 1--5~$\mu$m band. In the 1--2.5~$\mu$m near-infrared region, we identify the 1.0, 1.25, 1.6, and 2.1~$\mu$m flux peaks characteristic of cool brown dwarfs, shaped by strong H$_2$O and CH$_4$ bands \citep{2006ApJ...637.1067B}. This is followed by a broad minimum spanning 2.3--3.5~$\mu$m which encompasses H$_2$O, CH$_4$, and collision-induced H$_2$ absorption \citep{2015ApJ...799...37L,1969ApJ...156..989L}. A red peak emerges between 3.5 and 4.5~$\mu$m, showing the most structure for Source 39243. This region is shaped by overlapping bands of H$_2$O, CH$_4$, NH$_3$, CO$_2$, CO, and possibly PH$_3$ in cool brown dwarfs \citep{2010ApJ...722..682Y,2012ApJ...760..151S,2015ApJ...799...37L,2020AJ....160...63M,2023ApJ...951L..48B}, as discussed further below. We conclude from these spectral morphologies that all three sources are low-temperature brown dwarfs, and are hereafter designated as UNCOVER-BD sources.

\begin{deluxetable*}{lcccl}
\tabletypesize{\scriptsize}
\tablewidth{0pt} 
%\tablenum{1}
\tablecaption{Properties of the UNCOVER T Dwarfs \label{tab:properties}}
\tablehead{ %\\
\colhead{Property} & 
\colhead{UNCOVER-BD-1} & 
\colhead{UNCOVER-BD-2} & 
\colhead{UNCOVER-BD-3\tablenotemark{\scriptsize{a}}} &
\colhead{Ref} 
}
\startdata 
%\hline
%\hline
\multicolumn{5}{c}{HST \& JWST Astrometry \& Photometry\tablenotemark{\scriptsize{b}}} \\
\hline
Coordinate & J00140901-3022126 & J00141114-3021585 & J00140333-3021217 & [1] \\
MSA ID\tablenotemark{\scriptsize{c}} & 32265 & 33437 & 39243 & [1] \\
F814W (AB)	&	27.38$\pm$0.31	&		\nodata		&		\nodata		& [1] \\
F115W (AB)	&	28.26$\pm$0.11	&	27.34$\pm$0.06	&	28.00$\pm$0.06	& [1] \\
F125W (AB)	&		\nodata		&		\nodata		&	27.57$\pm$0.49	& [1] \\
F150W (AB)	&	28.06$\pm$0.09	&	28.10$\pm$0.10	&	28.81$\pm$0.18	& [1] \\
F160W (AB)	&		\nodata		&		\nodata		&	27.49$\pm$0.69	& [1] \\
F200W (AB)	&	28.04$\pm$0.10	&	28.87$\pm$0.19	&	29.71$\pm$0.35	& [1] \\
F277W (AB)	&	28.25$\pm$0.08	&	29.03$\pm$0.16	&	29.16$\pm$0.14	& [1] \\
F356W (AB)	&	27.57$\pm$0.03	&	28.06$\pm$0.06	&	27.69$\pm$0.03	& [1] \\
F410M (AB)	&	27.04$\pm$0.04	&	26.71$\pm$0.03	&		\nodata		& [1] \\
F444W (AB)	&	27.28$\pm$0.04	&	26.97$\pm$0.03	&	25.637$\pm$0.005	& [1] \\
\hline
\multicolumn{5}{c}{Spectral Properties} \\
\hline
{Integration Time (hr)} & 14.3 & 14.3 & 4.4 & [2] \\
Spectral Type & sdT1 & T6 & T8--T9 & [2] \\
 %& & & T8--T9 & [4] \\
T$_{eff}$\tablenotemark{\scriptsize{d}} (K) & 1300 (1100--1500) & 1000 (1000-1100) & 550 (500--700) & [2] \\
%& & & 650--700 & [4] \\
$\log{g}$\tablenotemark{\scriptsize{d}} (cgs) & 5.0 (4.5--5.0) & 5.25 (5.25-5.5) & 5.25 (5.0--5.5) & [2] \\
$[M/H]$\tablenotemark{\scriptsize{d}} & -1.0 & 0.0 & -0.5 & [2] \\
\hline
\multicolumn{5}{c}{Other Inferred Properties} \\
\hline
d$_{model}$\tablenotemark{\scriptsize{e}} (kpc) & 4.5$\pm$1.2 & 2.32$\pm$0.23 & 0.87$\pm$0.30 & [2] \\
%& & & 0.65$\pm$0.08 & [4] \\
%d$_{phot}$ (kpc) & TBD & TBD & TBD & [2] \\
%Z (kpc) & TBD & TBD & TBD & [2] \\
$\mathcal{P}$(thin disk) & $<$1\% & {16\%} & {75\%} & [2] \\
$\mathcal{P}$(thick disk) & {76\%} & {79\%} & {25\%} & [2] \\
$\mathcal{P}$(halo) & {24}\% & {6\%} & {$<$1\%} & [2] \\
\enddata
\tablenotetext{a}{aka GLASS-BD-1 {or ''Nonino's Dwarf''} \citep{2023ApJ...942L..29N}.}\vspace{-5pt}
\tablenotetext{b}{Photometry in AB magnitudes from \citet{2023arXiv230102671W} (internal version 3.0.0).}\vspace{-5pt}
% \tablenotetext{c}{Multi-Slit Array identification number in the catalog of \citet{2023arXiv230102671W}.}
\tablenotetext{c}{Multi-Slit Array identification number \citep[][internal version 2.2.1]{2023arXiv230102671W}.}\vspace{-5pt}
\tablenotetext{d}{Single values based on best fits to LOWZ models, while ranges for T$_{eff}$ and $\log{g}$ encapsulate the fits to all five models examined.}\vspace{-5pt}
\tablenotetext{e}{Based on scaling surface fluxes of best model fits to observed apparent spectral fluxes and assuming a radius of 1 Jupiter radius.}\vspace{-5pt}
%\tablenotetext{c}{Based on absolute magnitude relations [MORE]}
\tablerefs{
[1] \citet{2023arXiv230102671W};
[2] This paper.
%[4] \citet{2023ApJ...942L..29N}
}
%%%%%%
% 32265
% atmos 1300/5/0, logkzz=8, LC cloud
% btsettl 1100/5/-0.5 alpha=0.2
% karalidi 1300/4.5/0 logkzz=7
% lowz 1300/5/-1 logkzz=2 co=0.1 BEST
% sonora: 1500/5/0/co=0.5 WORST
%
% 33437
% atmos: 1100/5.5/0 LC cloud, chemeq
% btsettl 1000/5.5/0, alpha=0 WORST
% karalidi 1100/5.5/0 logkzz=4
% lowz 1000/5.25/0 logkzz=2 co=0.55 BEST
% sonora: 1100/5.5/0
%
% 39243
% atmos: 700/5.5/0 LC cloud chemeq WORST
% btsettl: 500/5.0/0 alpha=0
% karalidi: 700/5.5/0 logkzz=2
% lowz: 550/5.35/-0.5 logkzz=10 co=0.85 BEST
% sonora: 650/5.25/-0.5 co=1.0
%
\end{deluxetable*}

\section{Spectral Analysis}

\begin{figure*}[t]
\centering
\includegraphics[width=7cm]{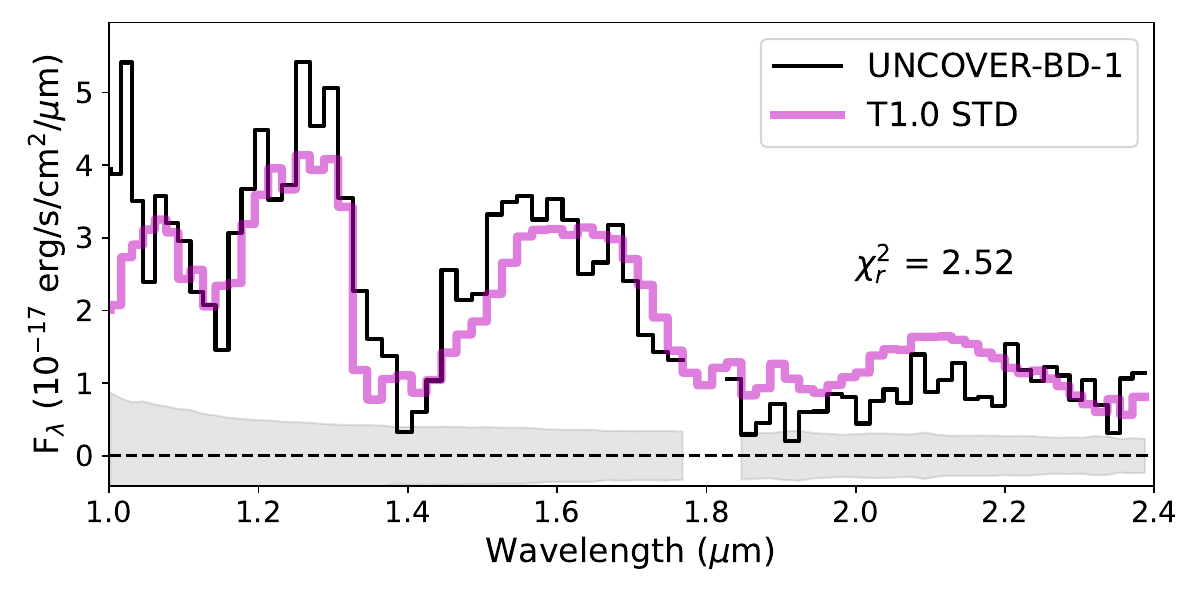}
\includegraphics[width=7cm]{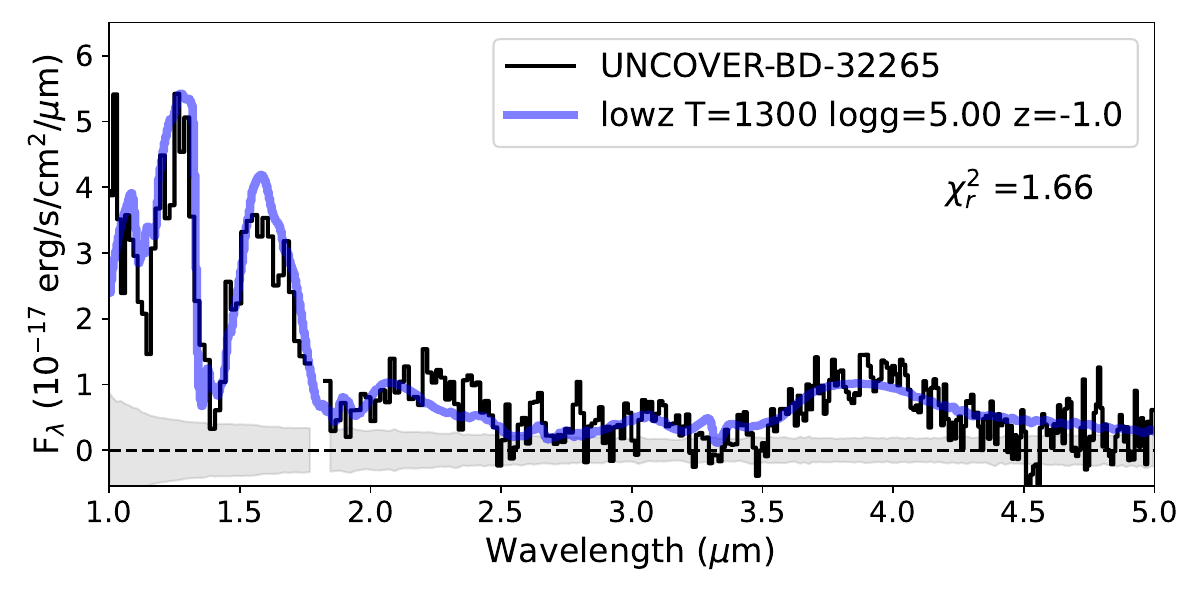} \\
\includegraphics[width=7cm]{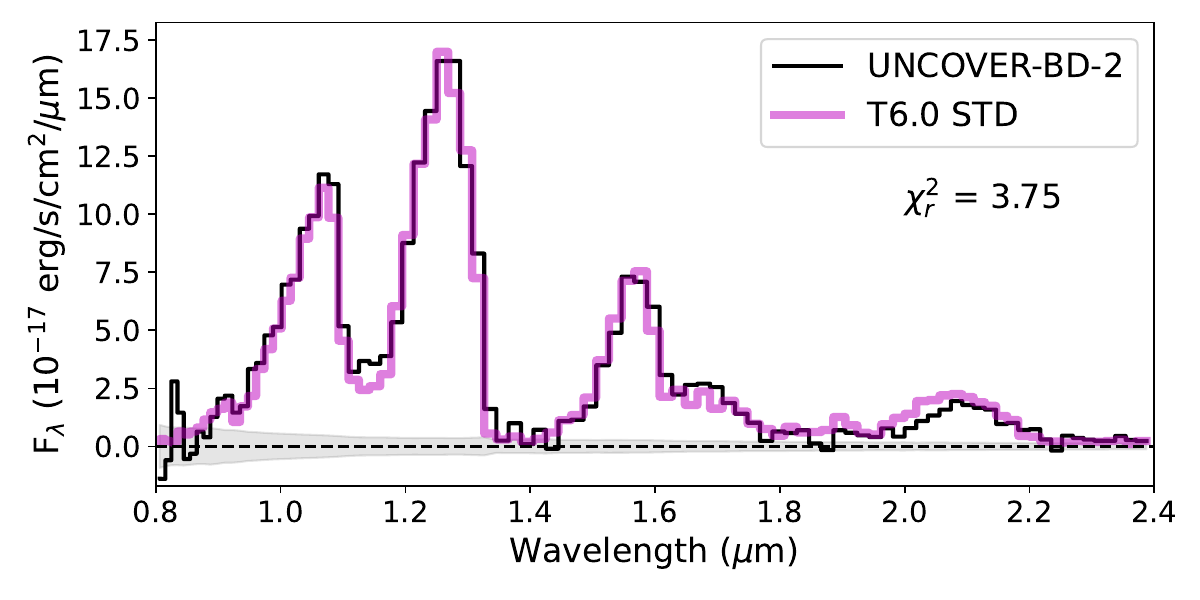}
\includegraphics[width=7cm]{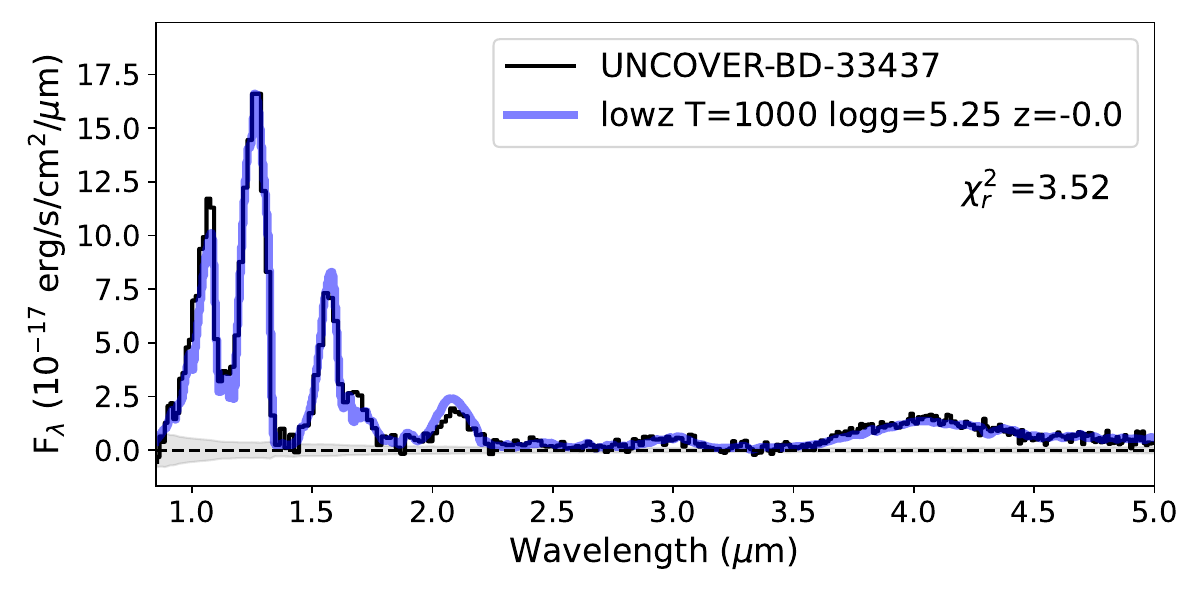} \\
\includegraphics[width=7cm]{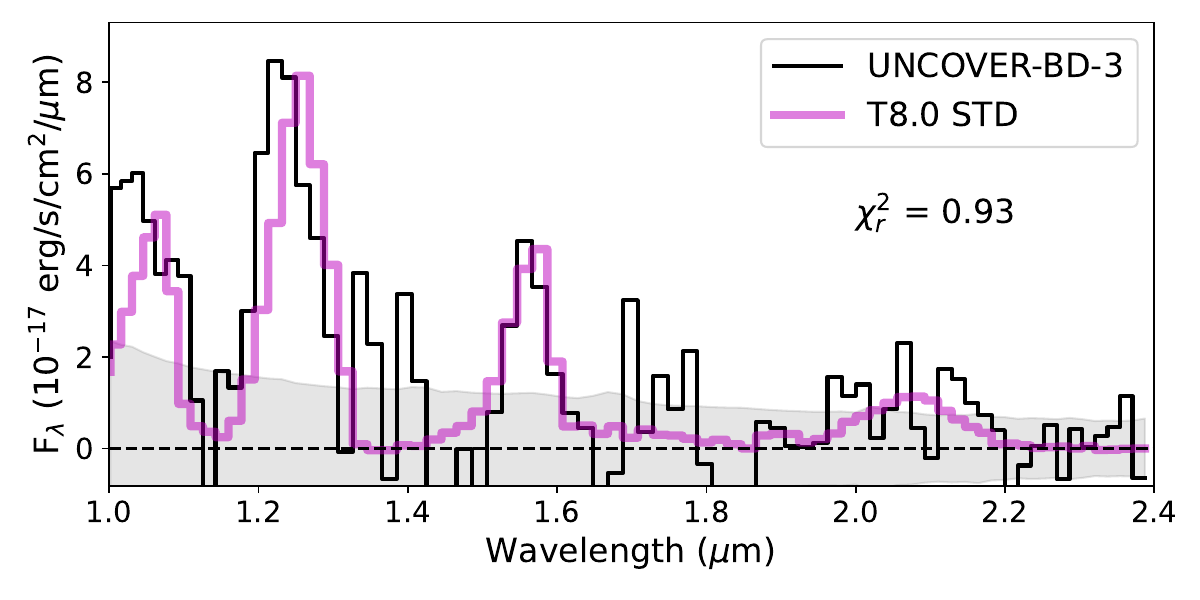}
\includegraphics[width=7cm]{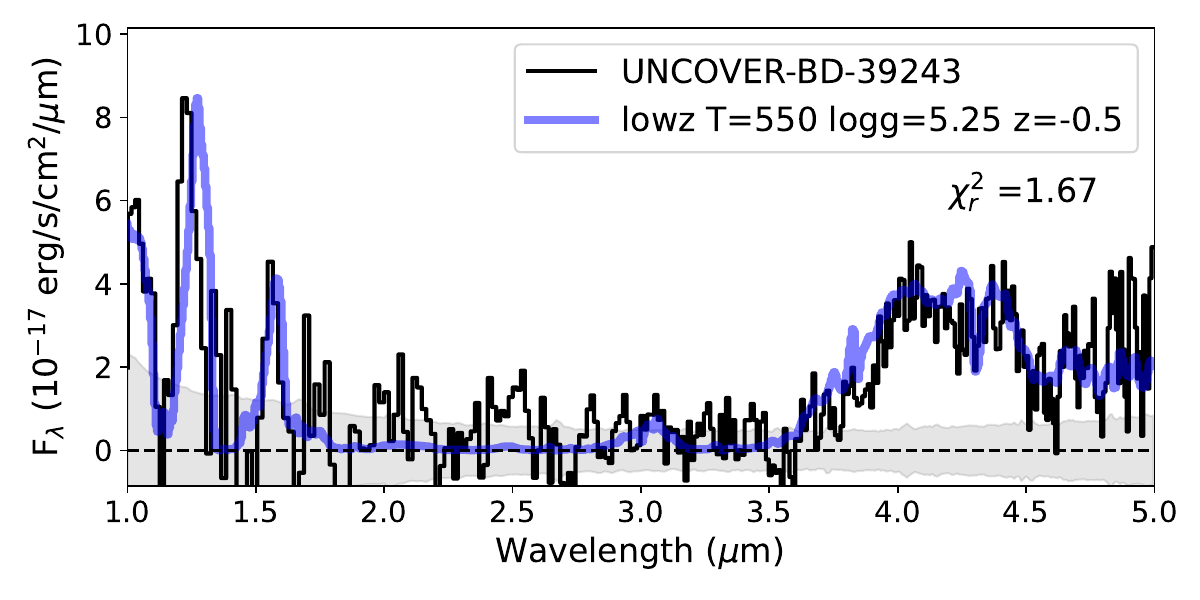} \\
\caption{(Left panels): Comparison of the 1-2.4~$\micron$ JWST/NIRSpec PRISM spectra (black lines with grey shaded uncertainties {in apparent flux density units}) to their best-fit near-infrared spectral standards from SPLAT (magenta lines). Standard spectra are from \citet{2004AJ....127.2856B,2006ApJ...639.1095B}. 
(Right panels): Best-fits LOWZ models to the full 1--5~$\micron$ spectra (blue lines), with parameters indicated in the legends.
\label{fig:analyze}}
\end{figure*}

We classified the sources by comparing their 1--2.4~$\mu$m spectra to L and T dwarf spectral standards from the SpeX Prism Library Analysis Toolkit (SPLAT; \citealt{2017ASInC..14....7B}). We identified the standard with the smallest reduced $\chi^2_r$ residual,\footnote{We compute chi-square as $\chi^2 \equiv \sum_i\frac{(O_i-\alpha{C_i})^2}{\sigma_i^2}$, where $\mathbf{O}$ is the observed spectrum, $\mathbf{C}$ the comparison template, $\mathbf{\sigma}$ the observed spectrum uncertainties, and $\alpha = \frac{\sum_iO_iC_i/\sigma_i^2}{\sum_iC_i^2/\sigma_i^2}$ is a scaling factor that minimizes $\chi^2$. We compute reduced chi-square as $\chi^2_r \equiv \chi^2/(N-1)$, where N is the number of spectral data points.} shown in the left panels of Figure~\ref{fig:analyze}. For UNCOVER-BD-1, the T1 standard provides the best overall fit, but has clear deviations in each of the $J$-, $H$-, and $K$-band spectral peaks. In particular, UNCOVER-BD-1 shows excess flux on the blue side of the $H$-band while having a relatively suppressed $K$-band. As discussed below, we infer these deviations are caused by subsolar metallicity. UNCOVER-BD-2 is an excellent match to the T6 standard, while UNCOVER-BD-3 equally matches T8 and T9 standards, {likely due to the spectrum's lower signal-to-noise.} 

To quantify the physical properties of these sources, we compared the JWST/NIRSpec spectra to five sets of low-temperature atmosphere models:
BT-Settl \citep{2012RSPTA.370.2765A},
{ATMO} \citep{2020A&A...637A..38P},
Sonora-Bobcat \citep{2021ApJ...920...85M},
Sonora-Cholla \citep{2021ApJ...923..269K}, 
and LOWZ \citep{2021ApJ...915..120M}.
These models span the temperature range of T- and Y-type dwarfs (roughly 500--1500~K), with varying parameters and treatments for metal composition, opacities, chemical equilibrium, and cloud formation. We focused on models with surface gravities $\log{g} \geq 4$ (cgs), consistent with ages $\gtrsim$100~Myr. 
We smoothed and interpolated each model to match the resolution of the NIRSpec/PRISM data over the 1--5~$\mu$m band, scaling model fluxes to minimize $\chi^2_r$ residuals. We performed a simple grid fit, identifying the lowest $\chi^2_r$ match to the data within each grid. 

Of the five model sets, the LOWZ models consistently provided the best fits (Figure~\ref{fig:analyze}), although all five models gave equivalent values for temperature and surface gravity. The LOWZ models are also computed at subsolar metallicities, and it is notable that both UNCOVER-BD-1 and UNCOVER-BD-3 are best fit to subsolar metallicity models. 
%Indeed, the parameters inferred for the former are similar to those inferred for two nearby metal-poor T subdwarf discoveries using the same models \citet{2020ApJ...898...77S}. 
Our inferred temperature for UNCOVER-BD-3, 500~K $\lesssim$ T$_{eff}$ $\lesssim$ 700~K, is also consistent with the 650~K estimate by \citet{2023ApJ...942L..29N} based on NIRCam photometry. Finally, we note that CO, CH$_4$, and NH$_3$ abundances are strongly affected by non-equilibrium mixing in brown dwarf atmospheres, and all three molecules exhibit absorption features in the 1--5~$\mu$m range (Figure~\ref{fig:spectra}). The best-fit ATMOS and Sonora-Cholla models both indicate enhanced vertical mixing with $\log\kappa_{zz}$ = 7--8 (cgs) for UNCOVER-BD-1, while UNCOVER-BD-2 and UNCOVER-BD-3 are best matched to low $\kappa_{zz}$/chemical equilibrium models. 

% \begin{figure*}[t]
% \centering
% \includegraphics[height=5.5cm]{magdist_32265.pdf}
% \includegraphics[height=5.5cm]{magdist_33437.pdf}
% \includegraphics[height=5.5cm]{magdist_39243.pdf}
% \caption{(TBD) Comparison of spectral model fit distance estimates (dashed line with blue band) and photometric/spectral type distance estimates (points with black error bars) for our three brown dwarfs. For the latter, we show the HST/JWST filter regions by horizontal magenta lines, and the filter regions of the absolute magnitude relations as horizontal black lines.  [WHY ARE THEY SO DIFFERENT?]
% \label{fig:distance}}
% \end{figure*}

The atmosphere models are computed as surface fluxes; hence, our fits {to apparent flux densities constraint the} scale factor $\alpha = (R/d)^2$, where $R$ is the radius of the source and $d$ its distance. Assuming a common radius of 1 Jupiter radius for all three brown dwarfs, averaging over all best-fit models, and ignoring the presence of binary companions, we infer the spectroscopic distance estimates listed in Table~\ref{tab:properties}.
Our distance estimate of 870$\pm$300~pc for UNCOVER-BD-3 overlaps with the 570-720~pc estimate by \citet{2023ApJ...942L..29N}, while the other two brown dwarfs have estimated distances exceeding 2~kpc. 
%We also estimated distances from HST and JWST photometry compiled in \citet{2023arXiv230102671W} and various infrared absolute magnitude/spectral type relations defined for local brown dwarfs.\footnote{We used relations from 'dupuy2012','dupuy2013','filippazzo2015','faherty2016','best2018','kirkpatrick2019','kirkpatrick2021','pecaut2013','tinney2014','zhang2019tsd'. In each case, we used the closest photometric band defined in these relations, and computed a color term by integrating that filter band and the HST/JWST band over the observed spectrum, appropriately accounting for Vega versus AB magnitudes.} [RESULTS]  
To our knowledge, these are the most distant T dwarfs to have measured spectroscopy.

\section{Population Constraints\label{sec:pop}}

As the Abell 2744 field is at a high Galactic latitude ($b$ = $-81\degr$), the kpc-scale distances of these brown dwarfs translate into large vertical offsets from the Galactic plane, and hence a high likelihood of being members of the thick disk or halo populations. 
To assess population membership {and to ascertain whether the number of T dwarfs found is consistent with expectations}, we constructed a Galactic population simulation of thin disk, thick disk, and halo low-mass stars and brown dwarfs, accounting for substellar evolution and population spatial distributions. 
{Full details of the simulation are provided in the Appendix.}

\begin{figure*}[t]
\centering
\includegraphics[height=5cm]{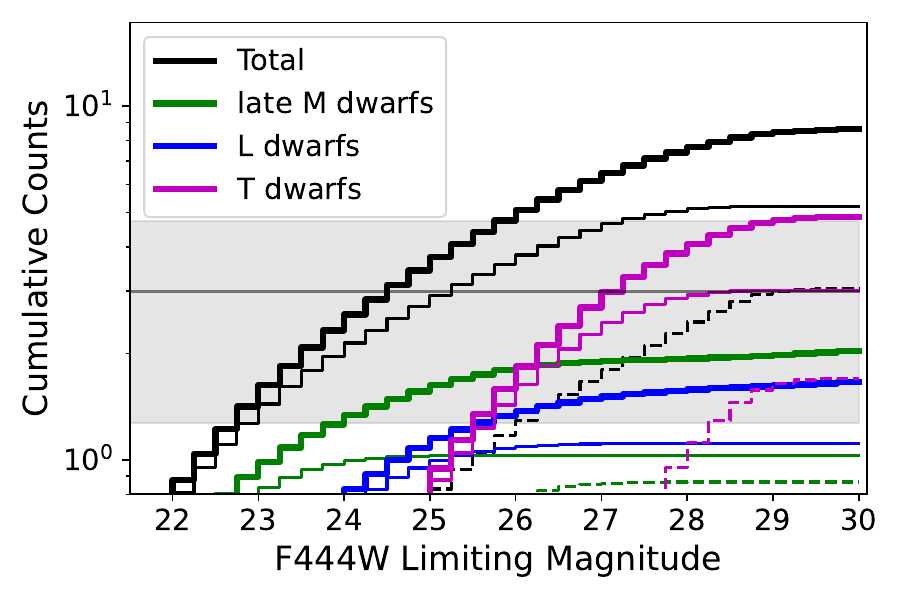}
\includegraphics[height=5cm]{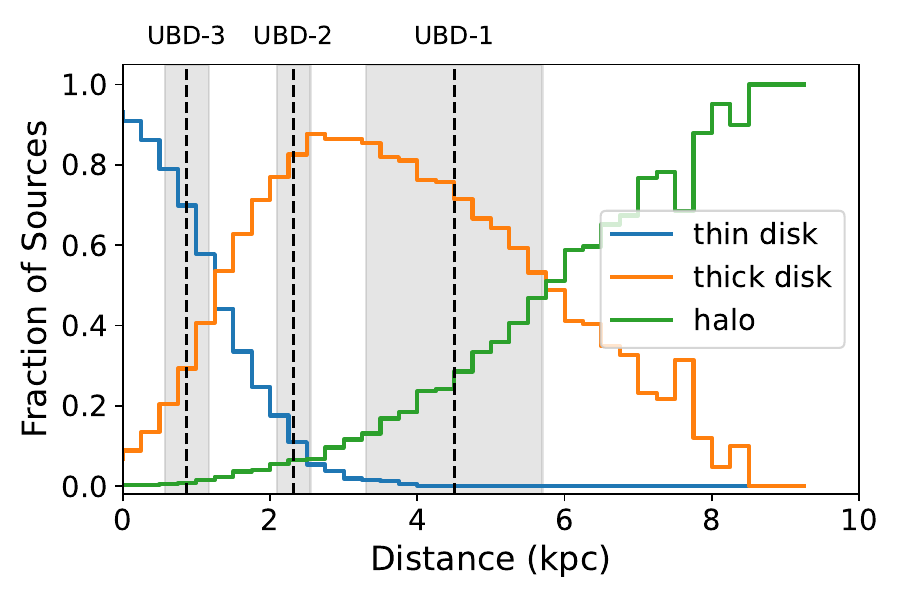}
\caption{(Left) Cumulative number counts of expected L and T dwarfs in the UNCOVER field of view as a function of limiting F444W magnitude. Sources are required to be detected ($<$30 AB mag) in at least three filters. Counts are broken down by spectral class (L dwarfs in blue, T dwarfs in magenta, total in black) and by population (thin disk as solid lines, thick disk as dashed lines, halo as dotted lines, total as thick solid lines). The grey horizontal line and band delineates the number of confirmed T dwarfs in the Abell 2744 field with Poisson uncertainties. 
(Right): Relative fraction of thin disk (blue line), thick disk (orange line), and halo (green line) as a function of distance for the T dwarfs in our simulation.
The vertical regions indicate the estimated distances and 1$\sigma$ uncertainties of the three T dwarfs reported here.
\label{fig:sim}}
\end{figure*}

Figure~\ref{fig:sim} shows the overall predicted numbers of detectable {ultracool dwarfs} in the Abell 2744 field based on these simulations, {with the numbers of late-M, L, and T dwarfs disaggregated. At magnitudes F444W $\gtrsim$ 26 AB}, T dwarfs are the dominant population of ultracool dwarfs {in this field}, and are primarily thin disk sources.  By F444W = 30 AB, our simulations predict {approximately 5} T dwarfs in the Abell 2744 field, {$\approx$60\% from the thin disk, $\approx$35\% from the thick disk, and $\lesssim$5\% from the halo}.
The detection of 3 T dwarfs in the limited spectroscopic follow-up of targets thus far, {and perhaps 1-2 strong candidates in photometric data \citep{2023arXiv230905714G},} is fully consistent with these simulations. 
{Only $\sim$2 M dwarfs and $\sim$1--2 L dwarfs are expected in the Abell 2744 field, in 1:1 and 2:1 ratios of thin:thick disk members; and a negligible number of Y dwarfs due to their intrinsic faintness. There are also few ($\sim$0-1) halo ultracool dwarfs expected in this field despite its depth, although this prediction is more sensitive to the poorly-constrained atmospheric and evolutionary properties of metal-poor brown dwarfs}. Hence, while the brown dwarf sample in the narrow Abell 2744 field may be {small}, it is a mixture of different Galactic populations that is distinct from the local Solar Neighborhood. 

Our simulations allow us to assess the individual population memberships of the three UNCOVER T dwarfs based on their estimated distances.  Figure~\ref{fig:sim} shows how the relative fractions of thin disk, thick disk, and halo T dwarfs varies as a function of distance along the Abell 2744 line of sight. Beyond {1.4~kpc}, thick disk {T} dwarfs start to outnumber thin disk {T} dwarfs; beyond {6~kpc}, halo {T} dwarfs start to outnumber thick disk {T} dwarfs. The estimated distances of the UNCOVER-BDs {span these thresholds, with UNCOVER-BD-1 in particular having 76\% and 24\% probabilities of being a thick disk or halo member, respectively.}
%; while UNCOVER-BD-3 has a 75\% probability of being a thin disk brown dwarf.}
Our small spectral sample is therefore representative of brown dwarfs in the Milky Way's three main populations. 

\section{Discussion of Individual Sources}

\subsection{The Distant T subdwarf UNCOVER-BD-1}

\begin{figure*}[t]
\centering
\includegraphics[height=5.5cm]{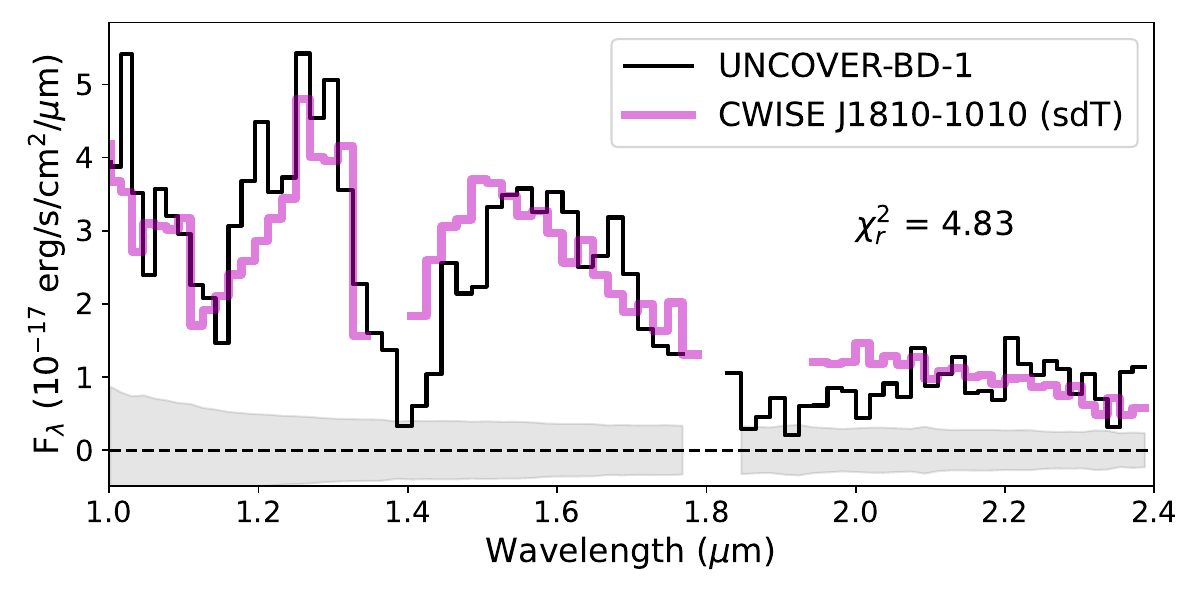}
\caption{Comparison of the spectra of UNCOVER-BD-1 (black line) and the extreme T subdwarf CWISE~J1810-1010 (magenta line; data from \citealt{2020ApJ...898...77S}). The latter is scaled to minimize the $\chi^2_r$ residuals between the data. Note that breaks in the CWISE~J1810-1010 spectrum represent regions of strong telluric absorption in ground-based spectroscopy.
\label{fig:sdT}}
\end{figure*}

The subsolar metallicity features in the spectrum of UNCOVER-BD-1 are fully consistent with this source's probable association with the thick disk or halo. The atmosphere parameters inferred for this source closely match those of CWISE~J1810-1010, a nearby T subdwarf identified in the citizen science Backyard Worlds: Planet 9 program \citep{2020ApJ...898...77S,2017ApJ...841L..19K}. We compare the spectrum of these two sources in Figure~\ref{fig:sdT}. The T subdwarf more accurately reproduces the widened 1.1~$\mu$m absorption feature and relatively featureless $H$-band peak characteristic of metal-poor brown dwarfs, but shows excess flux on the blue wings of the $H$- and $K$-band peaks. 
Of course, it is unlikely given the few T subdwarfs currently known that a perfect match would have occurred \citep{2019ApJS..240...31Z,2021ApJ...915..120M}.
Nevertheless, the similarities combined with both model fitting and population analysis supports the interpretation of this source as a metal-poor brown dwarf.
We note that parallax measurement of CWISE~J1810-1010 by \citet{2022A&A...663A..84L} has shown this source to be considerably cooler ($\sim$800~K) and closer (9~pc) then initial estimates, and the same may be true of our estimates of UNCOVER-BD-1. 

\subsection{Phosphine in the Spectrum of UNCOVER-BD-3?}

\begin{figure}[t]
\centering
\epsscale{1.2} 
\plotone{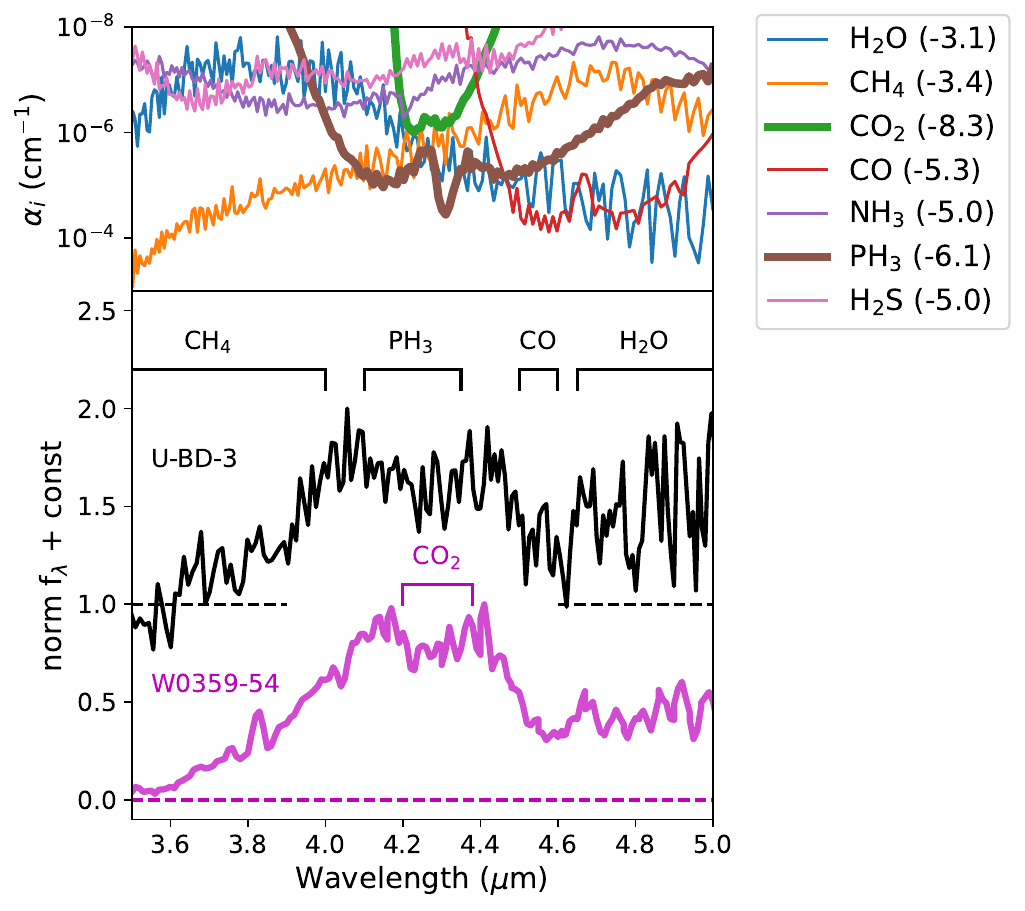}
\caption{(Top panel) Absorption coefficients for H$_2$O, CH$_4$, NH$_3$, CO$_2$, CO, PH$_3$, and H$_2$S in the 3.5--5~$\micron$ region based on EXOMOL cross-sections \citep{2013Icar..226.1673H,2012MNRAS.425...21T} and relative abundances based on equilibrium and non-equilibrium chemistry (log values of relative abundance for each species are listed in the legend; see text for details). The strongest absorber at a given wavelength is toward the bottom of this plot.
Absorption coefficents for CO$_2$ (in green) and PH$_3$ (in brown) are highlighted.
(Bottom panel) Normalized spectra of UNCOVER-BD-3 (black line ) and the Y0 dwarf W0359-54 (magenta line) in the 3.5--5~$\micron$ region, with the former offset for comparison (dashed lines). We identify the dominant absorbing species across this band based on the top panel, including PH$_3$ for UNCOVER-BD-3 and CO$_2$ for W0359-54.
\label{fig:ph3}}
\end{figure}

Our coldest source, UNCOVER-BD-3/GLASS-BD-1/{Nonino's Dwarf} is notable for the significant structure present in its 4--4.5~$\mu$m peak (Figure~\ref{fig:ph3}). This region is shaped by multiple molecular species, including H$_2$O, CH$_4$, NH$_3$, CO$_2$, CO, PH$_3$, and H$_2$S, with relative abundances influenced by non-equilibrium chemistry and metallicity, \citep{1999ApJ...519L..85G,2000ApJ...541L..75N,2006ApJ...647..552S,2006ApJ...648.1181V,2010ApJ...722..682Y,2015ApJ...799...37L,2020AJ....160...63M,2023ApJ...951L..48B}.
We compare the spectrum of UNCOVER-BD-3 in this region to the JWST/NIRSpec spectrum of the T$_{eff}$ $\approx$ 450~K Y0 dwarf WISE~J035934.06-540154.6 (hereafter W0359-54; \citealt{2012ApJ...753..156K,2016ApJ...824....2L,2023ApJ...951L..48B}). Both sources show absorption from the red wing of the 3.3~$\micron$ CH$_4$ $\nu_3$ band, the 4.5~$\micron$ bandhead of CO (enhanced by mixing), and H$_2$O extending toward and beyond 5~$\micron$. However, while
W0359-54 shows a distinct minimum over 4.2--4.4~$\micron$, a feature that has been associated with CO$_2$ absorption \citep{2010ApJ...722..682Y,2012ApJ...760..151S,2023ApJ...951L..48B}, this depression is broader and lacks the sharp CO$_2$ bandhead {in the spectrum of for UNCOVER-BD-3}. 

We investigated this distinction by comparing absorption coefficients $\alpha_i = n_i\sigma_i$ for the seven species $i$ listed above, using cross-sections $\sigma_i$ at T = 600~K provided by the EXOMOL cross-section server\footnote{See
\url{https://www.exomol.com/data/data-types/xsec}. Cross-sections are based on molecular line lists from \citet{2013Icar..226.1673H,2015ApJS..216...15L,2011MNRAS.413.1828Y,2018MNRAS.480.2597P,2016MNRAS.460.4063A,2015MNRAS.446.2337S}\ and \citet{2020MNRAS.496.5282Y}.} \citep{2013Icar..226.1673H,2012MNRAS.425...21T}. Relative abundances $n_i/n$ at T = 600~K and P = 10~bar ($n$ = 1.2$\times$10$^{20}$~cm$^{-3}$) are based on non-equilibrium chemistry for H$_2$O, CH$_4$, CO, NH$_3$ and PH$_3$ \citep{2006ApJ...647..552S,2006ApJ...648.1181V}, and equilibrium chemistry for H$_2$S \citep{2006ApJ...648.1181V}. For CO$_2$, we adopted the conjecture of \citet{2010ApJ...722..682Y} that N$_{CO2}$/N$_{CO}$ $\approx$ 10$^{-3}$, and thus exceeds equilibrium abundances by several orders of magnitude \citep{2002Icar..155..393L}.
\citet{2010ApJ...722..682Y} notes that this CO$_2$ enrichment {even exceeds expectations for disequilibrium mixing abundances, but} is necessary to explain the distinct bandhead seen in solar-metallicity T {and Y} dwarf spectra at 3--5~$\micron$. 
The mechanism for this enrichment has yet to be determined.  
%This high abundance is necessary to explain the distinct bandhead present in the 3--5~$\micron$ spectra of solar-metallicity T dwarfs \citet{2010ApJ...722..682Y,2012ApJ...760..151S}. 
Indeed, Figure~\ref{fig:ph3} shows that even with these assumptions, it is PH$_3$, not CO$_2$ that {should be} the primary absorber around 4.2~$\micron$. 
{Introducting into this analysis} reduced metal abundances in the atmosphere of UNCOVER-BD-3, suggested by our model fits and {modest} probability of thick-disk membership {(25\%)}, {would further} reduce the abundance of CO$_2$ relative to single-metal species such as H$_2$O or PH$_3$, just as TiO is reduced relative to CaH in the optical spectra of M-type subdwarfs \citep{1978ApJ...220..935M,1997AJ....113..806G}.
While the blue wing of the 4.2~$\micron$ feature in UNCOVER-BD-3 could be caused by a relative strengthening of H$_2$O over CH$_4$, we would expect to see more absorption in the 4.8--5.0~$\micron$ region as well, but instead see a slight rise in flux. Neither NH$_3$ nor H$_2$S show opacity trends that could explain the structure of this feature.

We therefore claim that the 4.2~$\micron$ feature in UNCOVER-BD-3 is PH$_3$, not CO$_2$, and its presence may serve as an indicator of subsolar metallicity among the coldest brown dwarfs.
The presence of multiple overlapping molecular species in this region, whose abundances are sensitive to mixing and metallicity, highlights its importance in atmospheric chemical studies of cold brown dwarfs and warm exoplanets \citep{2023ApJ...946L...6M}.

\section{Summary}

We report JWST/NIRSpec spectroscopy of three distant T dwarfs identified in the Abell 2744 lensing field by the UNCOVER JWST Legacy Survey. All three objects display the characteristic signatures of low temperature brown dwarf atmospheres, and are all well-matched in the 1--2.5~$\micron$ region to T1, T6, and T8--T9 standards. Spectral model fits confirm these identifications, and further indicate that UNCOVER-BD-1 and perhaps UNCOVER-BD-3 are metal-poor. These fits provide robust estimates of the distances of these sources that place them between 0.9--4.5~kpc, making them the most distant T dwarfs with confirming spectroscopy to date. {UNCOVER-BD-1 and UNCOVER-BD-2 in particular are sufficiently distant to have a} high probability of membership in the thick disk. 
Similarities in the spectra of UNCOVER-BD-1 and the T subdwarf CWISE~J1810-1010 supports the metal-poor nature of both sources. For UNCOVER-BD-3, the structure of the 4.2~$\micron$ peak indicates the signature of PH$_3$ absorption rather than CO$_2$ as seen in other T and Y dwarfs, and may reflect metallicity effects on molecular chemistry in cold brown dwarf atmospheres.
Our population simulations indicate that the Abell 2744 field should contain roughly {1--2} L-type and {5} T-type brown dwarfs, with {one-third of these} being members of the Galactic {thick disk}. While modest, this reflects the yield from a very small field of view (only seven NIRSpec pointings in this study), 
{and further investigations of an assemble of deep JWST pointings will uncover a larger and statistically useful sample for characterizing the oldest, metal-poor brown dwarfs in the Milky Way \citep{2023ApJ...942L..29N,2023arXiv230903250H,2023arXiv230905835H}.}

\begin{acknowledgments}
%\section*{}\noindent
%TBD - funding statements
{We dedicate this paper to the memory of Dr.\ Mario Nonino, discoverer of the first brown dwarf discovery made with JWST, GLASS-BD-1/UNCOVER-BD-3/Nonino's Dwarf, who passed way suddenly during the completion of this paper. Dr.\ Nonino will be remembered as a friendly and collaborative colleague.}
This work is based in part on observations made with the NASA/ESA \emph{Hubble Space Telescope} (HST) and the NASA/ESA/CSA \emph{James Webb Space Telescope} (JWST). 
Some of the data presented in this paper were obtained from the Mikulski Archive for Space Telescopes at the Space Telescope Science Institute.
%which is operated by the Association of Universities for Research in Astronomy, Inc., under NASA contract NAS 5-03127 for JWST, and NAS 5–26555 for HST. 
{The specific observations analyzed can be accessed via \dataset[10.17909/8k5c-xr27]{\doi{10.17909/8k5c-xr27}}}.
These observations are associated with JWST programs JWST-GO-2561, JWST-ERS-1324, and JWST-DD-2756; and with HST programs HST-GO-11689, HST-GO-13386, HST-GO/DD-13495, HST-GO-13389, HST-GO-15117, and HST-GO/DD-17231. Support for program JWST-GO-2561 was provided by NASA through a grant from the Space Telescope Science Institute, which is operated by the Associations of Universities for Research in Astronomy, Incorporated, under NASA contract NAS 5-03127. 
A.J.B. and R.G. acknowledge support from JWST program GO-2559
and NASA/ADAP grant 21-ADAP21-0187.
J.E.G. and A.D.G acknowledge support from NSF/AAG grant \# 1007094. 
J.E.G. also acknowledges support from NSF/AAG grant \# 1007052. 
A.Z. acknowledges support by Grant No. 2020750 from the United States-Israel Binational Science Foundation (BSF) and Grant No. 2109066 from the United States National Science Foundation (NSF), and by the
Ministry of Science \& Technology of Israel. The Cosmic Dawn Center is funded by the Danish National Research Foundation (DNRF) under grant \#140. This work has
received funding from the Swiss State Secretariat for Education, Research and Innovation (SERI) under contract number MB22.00072, as well as from the Swiss National Science Foundation (SNSF) through project grant 200020 207349. 
R.B. acknowledges support from the Research Corporation for Scientific Advancement (RCSA) Cottrell Scholar Award ID No: 27587 and from the National Science Foundation NSF-CAREER grant \# 2144314. 
P.D. acknowledges support from the Dutch Research Council (NWO) through the award of the VIDI Grant 016.VIDI.189.162 (``ODIN") and the European Commission's and University of Groningen's CO-FUND Rosalind Franklin program. 
R.P.N. acknowledges funding from JWST programs GO-1933 and GO-2279. Support for this work was provided by NASA
through the NASA Hubble Fellowship grant HST-HF2-51515.001-A awarded by the Space Telescope Science Institute, which is operated by the Association of Universities for Research in Astronomy, Incorporated, under NASA contract NAS5-26555. 
The research of C.C.W.
is supported by NOIRLab, which is managed by the
Association of Universities for Research in Astronomy
(AURA) under a cooperative agreement with the National Science Foundation.
R.P. and D.M. acknowledge support from JWST program GO-2561.

\end{acknowledgments}

%\vspace{5mm}
\facilities{JWST(NIRCam), JWST(NIRSpec), 
HST(ACS), HST(WFC3)}

\software{astropy \citep{2013A&A...558A..33A,2018AJ....156..123A,2022ApJ...935..167A},  
          msaexp \citep{Brammer2022},
          JWST Calibration pipeline \citep{bushouse_2023_8157276},
          SPLAT \citep{2017ASInC..14....7B}, 
          }

\appendix

\section{Ultracool Dwarf Galactic Population Simulations}

{Here we describe in detail the ultracool dwarf population simulations discussed in Section~\ref{sec:pop}. These simulations are based on Monte Carlo approaches previously presented in \citet{2004ApJS..155..191B,2007ApJ...659..655B,2019ApJ...883..205B,2021ApJS..257...45H}; and \citet{2022ApJ...934...73A}, with additional modifications described here to account for the spatial and spectral properties of ultracool dwarfs in different Galactic populations. See \citet{1999ApJ...521..613R,2013MNRAS.430.1171D,2013MNRAS.433..457B,2016AJ....151...92R,2016MNRAS.458..425V,2020AJ....159..257B,2021ApJS..253....7K}; and \citet{2022ApJ...932...96R} for other approaches.} 

{We modeled our ultracool dwarf sample as three populations representing the thin disk, the thick disk, and the halo, the latter treated as a single population (cf.\ \citealt{2008Natur.451..216C}). 
We generated a set of 5$\times$10$^5$ masses between 0.01-0.1~M$_\odot$ distributed as a power-law mass function with $dN/dM \propto M^{-0.6}$ \citep{2021ApJS..253....7K} that were common to the three populations; and three uniform distributions of ages, one spanning 0.5-8~Gyr for the thin disk, and two spanning 8--10~Gyr for the thick disk and halo. 
These samples were then evolved to present-day temperatures ($T_{eff}$), surface gravities ($\log{g}$), and luminosities using the Sonora-Bobcat models \citep{2021ApJ...920...85M}, applying the [M/H] = 0 models for the thin disk and [M/H] = $-$0.5 models for the thick disk and halo.\footnote{A comprehensive set of lower metallicity evolutionary models for cold brown dwarfs using up-to-date opacities are not yet publicly available; see \citet{2020RNAAS...4..214G}.}}

\begin{figure*}[t]
\centering
\plotone{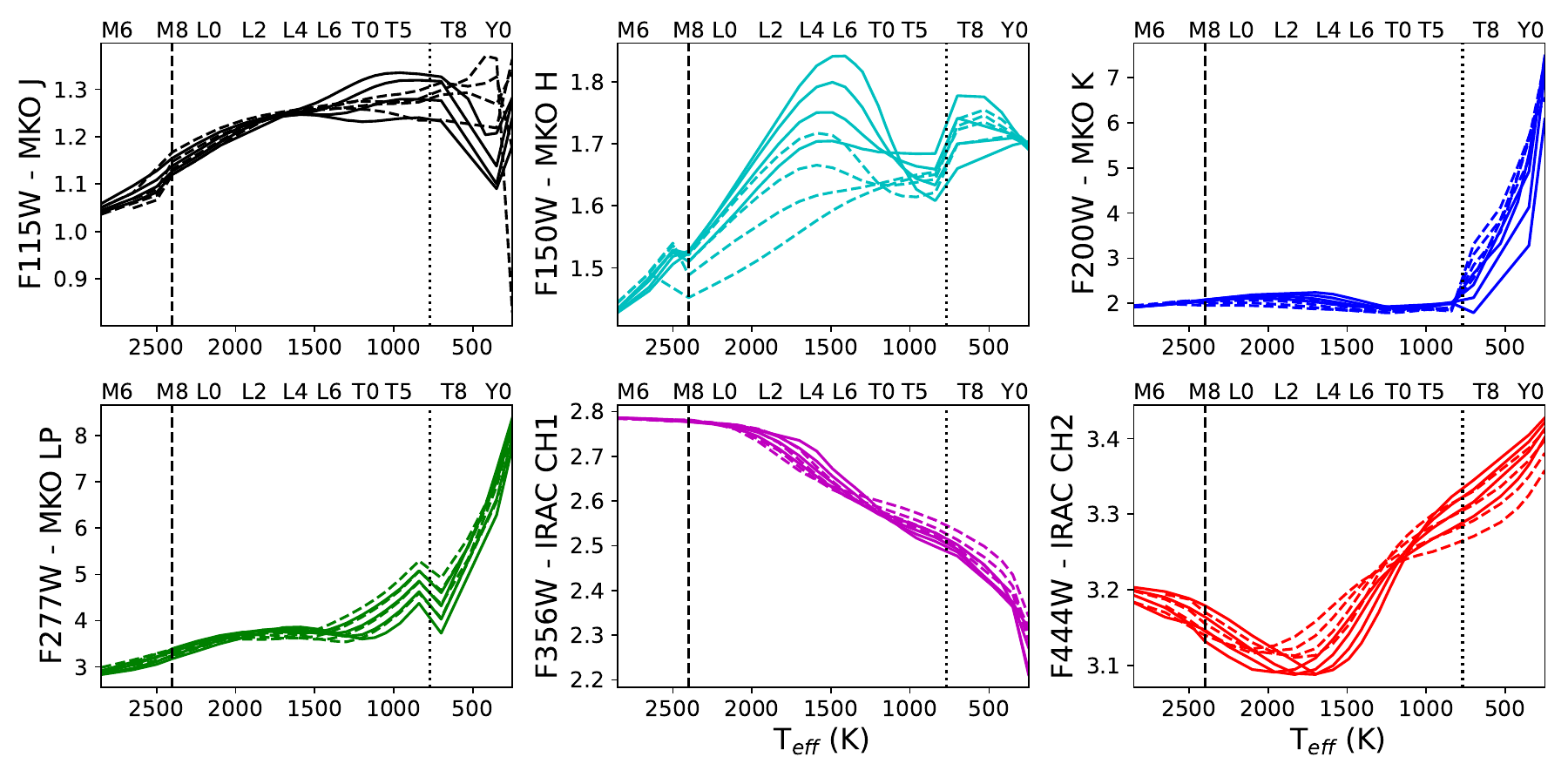}
\caption{{Filter color correction terms between JWST/NIRCam filters and reference filters used for empirical relations as a function of T$_{eff}$. Corrections for a variety of surface gravities and for solar-scaled metallicities [M/H] = 0 and -0.5 are shown by the various lines. Note that the models used to compute these corrections changes from BT-Settl \citep{2012RSPTA.370.2765A} to Sonora-Bobcat \citep{2021ApJ...920...85M} at T$_{eff}$ = 2400~K, and the reference filters and absolute magnitude relations used change at spectral type T8 (T$_{eff}$ = 700~K; see Table~\ref{tab:mags}).}
\label{fig:filtcolor}}
\end{figure*}

\begin{figure*}[t]
\centering
\plotone{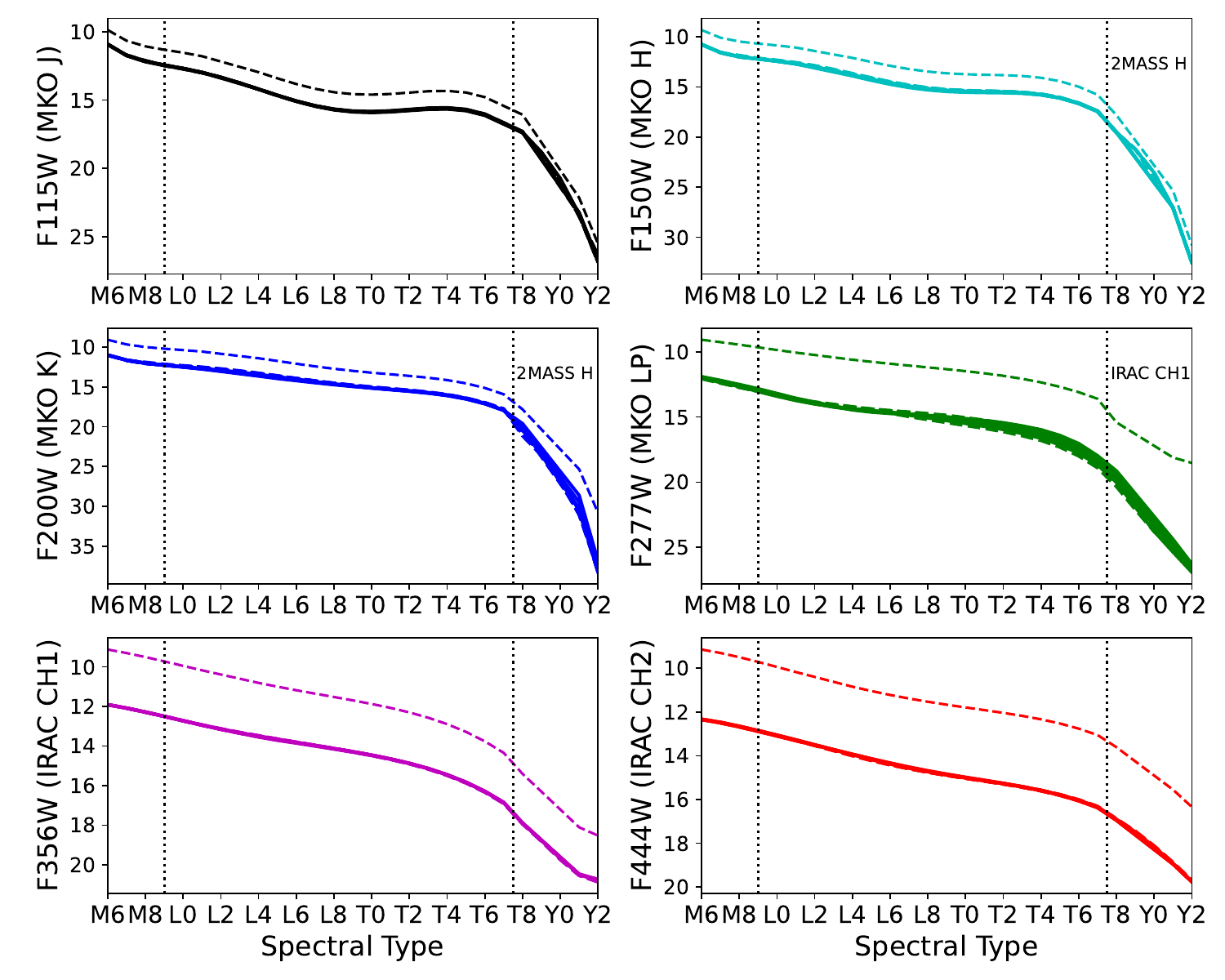}
\caption{{Absolute magnitudes in six JWST/NIRCam filters as a function of spectral type adopted in this study. The solid lines delineate the absolute magnitudes in the NIRCam filters in AB magnitudes, while the dashed lines indicate the original absolute magnitudes in the matched reference filter (indicated in the y-axis label and Table~\ref{tab:mags}). Note that the reference filter and empirical relation used changes at spectral type T7/T8. The increasing divergence between reference filter and NIRCam filter relations toward longer wavelengths largely reflects the divergence between the Vega and AB magnitude systems.}
\label{fig:absmagrelations}}
\end{figure*}

{These physical parameters were mapped to observational parameters of spectral type and absolute magnitude using a combination of empirical relations and theoretical atmosphere models. 
The empirical relations are based on a sample of local, largely solar-metallicity ultracool dwarfs, and are most appropriate for thin disk objects; see \citet{2019MNRAS.486.1260Z} and \citet{2021ApJ...923...19G} for relations appropriate for metal-poor M- and L-type ultracool dwarfs.
Temperatures were first mapped to spectral types from M6 to Y2 using an updated calibration from\footnote{E.\ Mamajek 2022; see \url{https://www.pas.rochester.edu/~emamajek/EEM_dwarf_UBVIJHK_colors_Teff.txt}.} \citet{2013ApJS..208....9P}. 
We then constructed spectral type/absolute magnitude relations for six wide-field JWST/NIRCam filters, F115W, F150W, F200W, F277W, F356W, and F444W,
anchoring to existing spectral type/absolute magnitude relations in proximate filters. 
For spectral types M6 to T7, we used the spectral type/absolute magnitude relations of \citet{2012ApJS..201...19D} for filters MKO~J, MKO~H, MKO~K, MKO~L$_P$, IRAC [3.6], and IRAC [4.5]. 
For spectral types T8 to Y2, we used the spectral type/absolute magnitude relations of \citet{2021ApJS..253....7K} for filters MKO~J, 2MASS~H, IRAC [3.6], and IRAC [4.5]. 
Color terms between these filters and JWST/NIRCam filters were computed using atmosphere models of the appropriate temperature and metallicity, and for surface gravities 4.0 $\leq$ {\logg} $\leq$ 5.5 (in cm/s$^2$). For T$_{eff}$ $>$ 2400~K, we used the BT-Settl models \citep{2012RSPTA.370.2765A}; for T$_{eff}$ $\leq$ 2400~K we used the Sonora-Bobcat models \citep{2021ApJ...920...85M}. 
Figure~\ref{fig:filtcolor} displays the color terms between the empirical and JWST/NIRCam filters, and Figure~\ref{fig:absmagrelations} displays the resulting JWST/NIRCam absolute magnitude/spectral type relations, tabulated in Table~\ref{tab:mags}.
Uncertainties on these values vary as a function of spectral type, and a conservative estimate of 0.5~mag is appropriate for most of the spectral type range shown. Comparable relations for a broader set of JWST/NIRCam filters can be found in \citet{2023RNAAS...7..194S}.}

{Distances for each source were assigned by drawing from projected stellar density distributions along the Abell 2744 line of sight using the Milky Way structure parameters of \citet{2008ApJ...673..864J}. The maximum distances for these distributions were set by the absolute F444W magnitude of each source and an assumed limiting magnitude of 30 AB. 
For the M dwarfs in the simulation, maximum distances can exceed 30~kpc; for the T dwarfs, maximum distances range over 3--10~kpc.
We ignored both unresolved multiplicity and statistical scatter for these limits, factors that were considered insignificant given the small size of our search area and final sample. 
Each source was assigned a distance according to the line-of-sight density distribution for its population truncated at this maximum distance, and then assigned a corresponding apparent magnitude.} 

{To construct a composite population, we randomly drew sources with replacement from our thin disk, thick disk, and halo samples, ensuring that the relative numbers of sources within 500~pc matched the local fractions in the \citet{2008ApJ...673..864J} model: thick disk/thin disk = 12\% and halo/thin disk = 0.5\%. We further required that apparent magnitudes in at least three NIRCam filters were brighter than the 30 AB limit to ensure multi-band detection. Finally, we computed a scaling factor\footnote{The scaling factor $\frac{N_{\rm true}}{N_{\rm sim}} = \frac{\rho_{obs}}{N_{\rm sim}(<d)}\frac{\Omega}{3}d^3$ assumes a uniform spatial volume within our scaling distance $d$ = 100~pc; see \citet{2019ApJS..240...19K} and \citet{2020AJ....159..257B} for discussion on the spatial isotropy of ultracool dwarfs within 20~pc of the Sun.} to correct the cumulative number of L1--T8 dwarfs within $d$ = 100~pc in our simulation to the expected number of these sources in the same volume based on the measured local space density $\rho_{obs}$ = (1.09$\pm$0.06)$\times$10$^{-3}$~pc$^{-3}$ \citep{2021ApJS..253....7K} and the imaged area of $\Omega$ = 45~arcmin$^2$.} 

\clearpage

\startlongtable
\begin{deluxetable*}{lcclccccccccccccl}
\tabletypesize{\scriptsize}
\tablewidth{0pt} 
%\tablenum{1}
\tablecaption{JWST/NIRCam Absolute Magnitude Relations \label{tab:mags}}
\tablehead{ %\\
\colhead{SpT} & 
\colhead{T$_{eff}$ (K)\tablenotemark{a}} & 
\colhead{[M/H]} & 
\colhead{Model} &
\colhead{F115W} &
\colhead{F150W} &
\colhead{F200W} &
\colhead{F277W} &
\colhead{F356W} &
\colhead{F444W} 
}
\startdata 
\multicolumn{4}{c}{Base magnitude M6-T7\tablenotemark{b}} & 
MKO J &
MKO H &
MKO K &
MKO L$_P$ &
IRAC [3.6] &
IRAC [4.5] \\
\hline
M6        &        2850&           0&  BT        &      10.918&      10.769&      10.985&      11.985&      11.911&      12.328\\
M6        &        2850&        -0.5&  BT        &      10.910&      10.779&      11.009&      12.055&      11.910&      12.329\\
M7        &        2650&           0&  BT        &      11.742&      11.585&      11.631&      12.354&      12.085&      12.467\\
M7        &        2650&        -0.5&  BT        &      11.727&      11.592&      11.645&      12.397&      12.084&      12.473\\
M8        &        2500&           0&  BT        &      12.168&      12.012&      12.022&      12.664&      12.285&      12.651\\
M8        &        2500&        -0.5&  BT        &      12.147&      12.023&      12.040&      12.726&      12.284&      12.656\\
M9        &        2400&           0&  SON       &      12.461&      12.205&      12.254&      13.044&      12.502&      12.843\\
M9        &        2400&        -0.5&  SON       &      12.474&      12.147&      12.149&      13.030&      12.506&      12.851\\
L0        &        2250&           0&  SON       &      12.706&      12.420&      12.458&      13.404&      12.722&      13.052\\
L0        &        2250&        -0.5&  SON       &      12.717&      12.342&      12.325&      13.362&      12.725&      13.068\\
L1        &        2100&           0&  SON       &      12.990&      12.675&      12.665&      13.732&      12.939&      13.270\\
L1        &        2100&        -0.5&  SON       &      12.998&      12.577&      12.506&      13.643&      12.936&      13.297\\
L2        &        1960&           0&  SON       &      13.338&      12.999&      12.892&      13.996&      13.141&      13.499\\
L2        &        1960&        -0.5&  SON       &      13.344&      12.882&      12.719&      13.846&      13.130&      13.534\\
L3        &        1830&           0&  SON       &      13.748&      13.384&      13.147&      14.198&      13.326&      13.727\\
L3        &        1830&        -0.5&  SON       &      13.754&      13.255&      12.974&      14.022&      13.309&      13.770\\
L4        &        1700&           0&  SON       &      14.201&      13.806&      13.421&      14.353&      13.494&      13.954\\
L4        &        1700&        -0.5&  SON       &      14.204&      13.672&      13.268&      14.214&      13.479&      14.002\\
L5        &        1590&           0&  SON       &      14.660&      14.224&      13.716&      14.496&      13.656&      14.175\\
L5        &        1590&        -0.5&  SON       &      14.659&      14.097&      13.600&      14.445&      13.648&      14.220\\
L6        &        1490&           0&  SON       &      15.090&      14.606&      14.026&      14.659&      13.812&      14.382\\
L6        &        1490&        -0.5&  SON       &      15.081&      14.494&      13.948&      14.706&      13.813&      14.419\\
L7        &        1410&           0&  SON       &      15.451&      14.928&      14.339&      14.855&      13.966&      14.568\\
L7        &        1410&        -0.5&  SON       &      15.434&      14.833&      14.285&      14.962&      13.977&      14.596\\
L8        &        1350&           0&  SON       &      15.714&      15.176&      14.637&      15.066&      14.124&      14.732\\
L8        &        1350&        -0.5&  SON       &      15.689&      15.092&      14.590&      15.198&      14.139&      14.751\\
L9        &        1300&           0&  SON       &      15.865&      15.342&      14.901&      15.273&      14.286&      14.880\\
L9        &        1300&        -0.5&  SON       &      15.834&      15.267&      14.857&      15.423&      14.305&      14.892\\
T0        &        1260&           0&  SON       &      15.908&      15.437&      15.134&      15.493&      14.461&      15.014\\
T0        &        1260&        -0.5&  SON       &      15.871&      15.369&      15.085&      15.647&      14.482&      15.022\\
T1        &        1230&           0&  SON       &      15.861&      15.485&      15.341&      15.712&      14.655&      15.142\\
T1        &        1230&        -0.5&  SON       &      15.821&      15.421&      15.289&      15.870&      14.678&      15.147\\
T2        &        1200&           0&  SON       &      15.765&      15.522&      15.545&      15.958&      14.878&      15.273\\
T2        &        1200&        -0.5&  SON       &      15.720&      15.463&      15.489&      16.119&      14.903&      15.276\\
T3        &        1160&           0&  SON       &      15.673&      15.603&      15.780&      16.269&      15.142&      15.417\\
T3        &        1160&        -0.5&  SON       &      15.623&      15.548&      15.715&      16.433&      15.167&      15.415\\
T4        &        1120&           0&  SON       &      15.650&      15.783&      16.084&      16.630&      15.459&      15.582\\
T4        &        1120&        -0.5&  SON       &      15.596&      15.732&      16.010&      16.796&      15.486&      15.577\\
T5        &        1050&           0&  SON       &      15.773&      16.122&      16.507&      17.131&      15.842&      15.782\\
T5        &        1050&        -0.5&  SON       &      15.713&      16.077&      16.416&      17.304&      15.870&      15.772\\
T6        &         960&           0&  SON       &      16.116&      16.672&      17.099&      17.780&      16.308&      16.028\\
T6        &         960&        -0.5&  SON       &      16.054&      16.634&      16.982&      17.987&      16.337&      16.011\\
T7        &         840&           0&  SON       &      16.754&      17.472&      17.907&      18.678&      16.874&      16.335\\
T7        &         840&        -0.5&  SON       &      16.702&      17.438&      17.747&      18.891&      16.902&      16.314\\
\hline
\multicolumn{4}{c}{Base magnitude T8-Y2\tablenotemark{c}} & 
MKO J &
2MASS H &
2MASS H &
IRAC [3.6] &
IRAC [3.6] &
IRAC [4.5] \\
\hline
T8        &         700&           0&  SON       &      17.393&      19.593&      20.479&      20.010&      17.909&      16.907\\
T8        &         700&        -0.5&  SON       &      17.365&      19.538&      21.128&      20.329&      17.940&      16.881\\
T9        &         530&           0&  SON       &      18.808&      21.234&      22.785&      21.909&      18.772&      17.483\\
T9        &         530&        -0.5&  SON       &      18.851&      21.193&      23.583&      22.095&      18.813&      17.449\\
Y0        &         420&           0&  SON       &      20.733&      23.561&      26.062&      23.767&      19.687&      18.159\\
Y0        &         420&        -0.5&  SON       &      20.899&      23.533&      26.853&      23.787&      19.738&      18.121\\
Y1        &         350&           0&  SON       &      23.362&      27.059&      30.632&      25.362&      20.476&      18.916\\
Y1        &         350&        -0.5&  SON       &      23.519&      27.048&      31.020&      25.201&      20.541&      18.876\\
Y2        &         250&           0&  SON       &      26.810&      32.485&      38.047&      26.911&      20.737&      19.743\\
Y2        &         250&        -0.5&  SON       &      26.356&      32.489&      37.352&      26.796&      20.868&      19.699\\
\enddata
\tablenotetext{a}{Temperatures estimated from the spectral type/T$_{eff}$ relation of \citet{2013ApJS..208....9P}.}
\tablenotetext{b}{Based on the \citet{2012ApJS..201...19D} absolute magnitude/spectral type relations.}
\tablenotetext{c}{Based on the \citet{2021ApJS..253....7K} absolute magnitude/spectral type relations.}
\tablecomments{Model references:
[BT] BT-Settl models \citep{2012RSPTA.370.2765A};
[SON] Sonora-Bobcat models \citep{2021ApJ...920...85M}.
}
\end{deluxetable*}

\clearpage

\bibliography{uncover_tdw}{}
\bibliographystyle{aasjournal}

\end{document}